\documentclass[12pt,
]{svjour3}
\pdfoutput=1
\usepackage{graphicx}
\usepackage{epstopdf}
\usepackage{amsfonts}
\usepackage{amssymb}
\usepackage{amsmath}
\usepackage{bm}
\usepackage{hyperref}
\usepackage{mathrsfs}%
\usepackage{bbm}
\usepackage{cancel}
\usepackage{subfigure}
\usepackage{color}
\usepackage{breqn}
\usepackage[numbers,sort&compress]{natbib}
\def\be{\begin{equation}}
\def\ee{\end{equation}}
\def\bea{\begin{eqnarray}}
\def\eea{\end{eqnarray}}

\begin{document}
\title{Following the primordial perturbations through a
bounce with AdS/CFT Correspondence}
\author{Lei~Ming, Taifan~Zheng, and Yeuk-Kwan~E.~Cheung}

\institute{ Lei~Ming, Taifan~Zheng, and Yeuk-Kwan~E.~Cheung
\at
Department of Physics, Nanjing University,
22 Hankou Road, Nanjing, China 210093}
\maketitle

\begin{abstract}
A bounce universe  model, known as the coupled-scalar-tachyon bounce (CSTB) universe, has been shown to  solve the Horizon, Flatness and Homogeneity problems as well as the Big Bang Singularity problem. Furthermore a  scale invariant spectrum of primordial density perturbations generated from the phase of pre-bounce contraction is shown to be stable against time evolution. In this work we study the detailed dynamics of the bounce and its imprints  on the scale invariance of the spectrum. The dynamics of the gravitational interactions near the bounce point may be strongly coupled as the spatial curvature becomes big. There is no a prior reason to expect the spectral index of the  primordial perturbations of matter density can be preserved. By encoding  the bounce dynamics  holographically onto the dynamics of dual Yang-Mills system while the latter is weakly coupled, via the AdS/CFT correspondence, we can safely evolve the spectrum of the cosmic perturbations  with full control. In this way we can compare the post-bounce spectrum with the pre-bounce one: in the CSTB model we explicitly show that the spectrum of primordial density perturbations generated in the contraction phase preserves its stability as well as scale invariance throughout the bounce process.
%
\end{abstract}

\section{Introduction}
A universe with a bounce process has emerged over the last decade  as a candidate theory for early universe dynamics as it can potentially produce a scale invariant spectrum of primordial density perturbations that   matched
up to  the current CMB measurements.
The key observation came from D. Wands when he noticed in 1998 that a scale invariant spectrum can be obtained in  a phase of matter dominated contraction prior to a
Big Bounce in his seminal work~\cite{Wands:1998yp}.
Although the spectrum followed from Wands's simple model
 was soon proven to be unstable~\cite{Gratton:2003pe},
 it opened up a new  arena in which alternatives to Inflation were
 explored~\footnote{The inflation scenario, and its many models, is reviewed in~\cite{Guth:2013sya, Linde:2014nna} in light of the Planck data~\cite{CMB}.}.
 Explicit model constructions ensue in the next two decades.
This active research area has been well covered in many nice reviews over the
years, the readers are referred
to~\cite{Nojiri:2017ncd, Brandenberger:2016vhg, Battefeld:2014uga, WilsonEwing:2012pu, Novello:PhysRept, Cai:2012va, Cai:2011zx, Lin:2010pf, Wang:2009rw, Piao:2009ku,
Peter:2008qz, Allen:2004vz, Lidsey:1999mc, Gasperini:1992em, Buonanno:1997zk,
Khoury:2001wf, Lyth:2001pf, Cai:2016hea, Addazi:2016rnz, Finelli:2002prd}
for a  repertoire of creative models, and further references.

In this article we shall continue our study of the coupled scalar-tachyon bounce
 (CSTB)  universe model  constructed from a  Rolling Tachyon  and
its coupling to the adjoint Higgs fields  in a system of non-BPS  D-branes and
anti-D-branes.
 In addition to the original model of Tachyon Inflation proposed, independently,
by Sen~\cite{Sen:2002in, Sen:2002nu} and  Gibbons~\cite{Gibbons:2002md}
  we incorporated the interactions of tachyon fields with the
 Higgs  fields   to model the  bounce process~\cite{Li:2011nj}.
The adjoint (real) Higgs fields live on the D-branes, and
they  encode,  geometrically,  the transverse separations  between the two stacks
of D-branes.  The CSTB model has been  shown to solve the Big Bang Singularity,
Horizon as well as  Flatness problems with generic values of the free
parameters~\cite{Li:2012vi, Cheung:2016oab}.
Furthermore a scale invariant
spectrum generated  during the phase of contraction is  shown to be stable;
it has no  implicit nor explicit time dependence  in the power spectrum and
it is thus stable against time evolution~\cite{Li:2013bha}.

\begin{figure}[h!]
\centering
\includegraphics[width=0.8\textwidth]{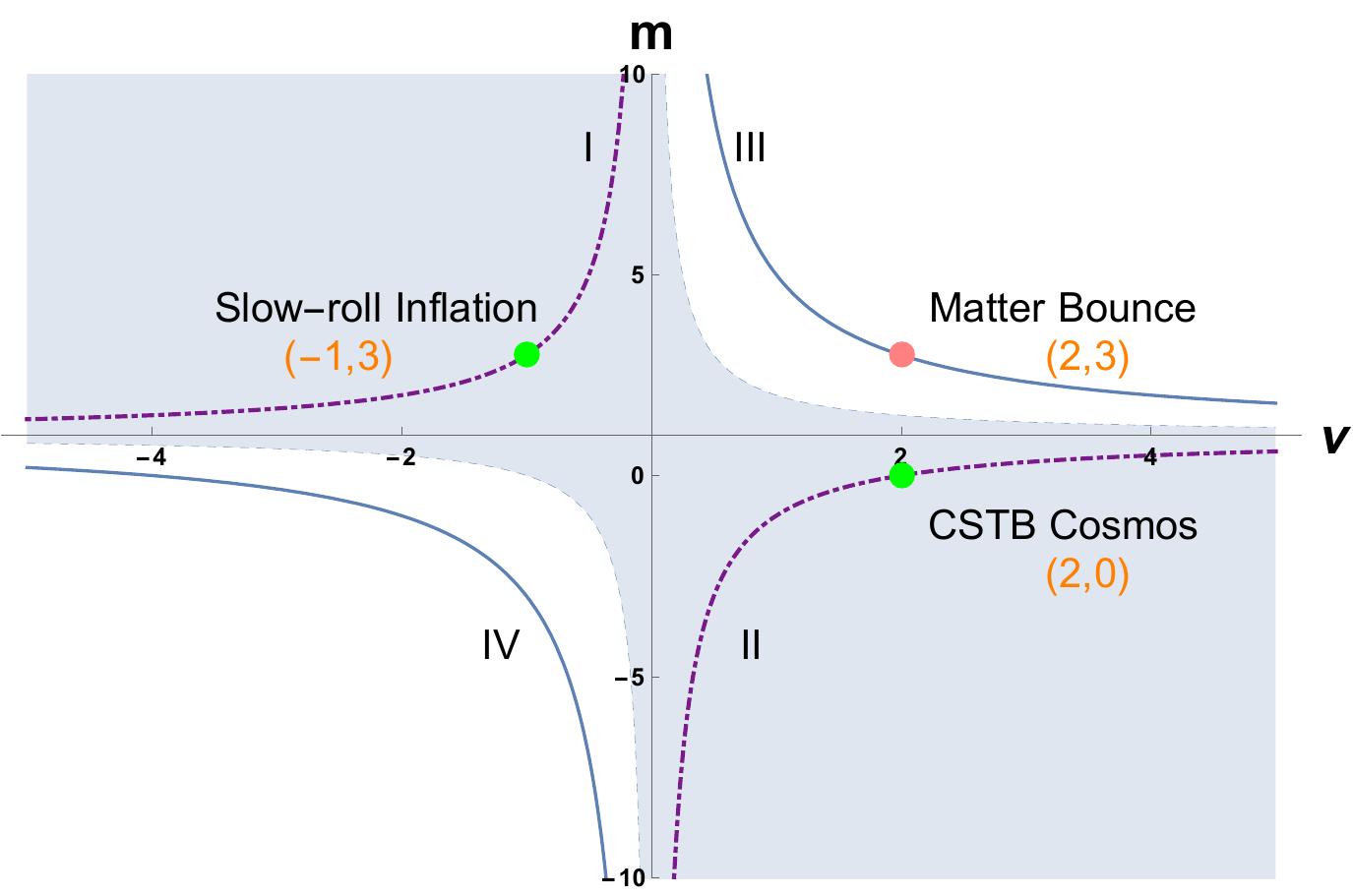}
\caption{%
The shaded region includes all time-independent solutions satisfying
$(m-1)\,\nu-1 < 0$  whose boundaries are defined by $(m-1)\,\nu-1= 0$
delineated by thin dash lines.
The purple dot-dash lines obeying $(m-1)\,\nu =-2$
represent scale-invariant as well as time-independent solutions.
Another set of scale invariant solutions given by
$(m-1)\,\nu =4 $ (violet solid lines) have Fourier modes varying with time and therefore are not truly scale-invariant in a physical sense.%
}\label{fig:duality}
\end{figure}

In~\cite{Li:2013bha} we also proposed a method to
analyse  early universe models--inflation or its alternatives--en masse
by noticing that the time independence, $\eta^{W(\nu,m)}$, and scale invariance,
$k^{L(\nu,m)}$,  are determined  by two relations,
\begin{equation}   \label{eq:L}
L(\nu,m) +3 =-|(m-1)\nu-1|+3=0,
\end{equation}
and
\begin{equation} \label{eq:W}
W(\nu,m)=[(m-1)\nu-1]+|(m-1)\nu-1|=0~,
\end{equation}
in the cosmic background evolved according to  $a^{\eta}$
and the Hubble term given by $mH$. 
For inflation models with a  canonical scalar  $m=3$,
with  $3$ coming  from the spatial dimensions of our presently observable universe.
In more general models $mH=3f(\phi_i,\dot{\phi}_i)H$
with $f$ being a generic function  determined by the underlying models. 
This method is applicable to  any early universe models in which
 the equation of the perturbation modes,  ${\chi}_k$,  in k-space obeys:
 \begin{equation} \label{eq:chi_k}
\ddot{\chi}_k+mH\dot{\chi}_k+\frac{k^2}{a^2}\chi_k=0~,\quad a\propto \eta^\nu~, \end{equation}
with $m(\eta, k)$  becoming  a general function of conformal time and  k-modes.

In Figure~\ref{fig:duality} general solutions to~(\ref{eq:L}) and~(\ref{eq:W})
are plotted in the  ($\nu$, m) parameter space:
the shaded region includes all time-independent solutions satisfying
$(m-1)\,\nu-1 < 0$  whose boundaries are defined by $(m-1)\,\nu-1= 0$
delineated by thin dash lines. The purple dot-dash lines,
 obeying $(m-1)\,\nu =-2$,  represent scale-invariant and  time-independent solutions. Another set of scale-invariant solutions given by
$(m-1)\,\nu =4 $ (blue solid lines) have  Fourier modes  varying with time and
therefore are not truly scale-invariant in a physical sense.
A few interesting solutions are marked in the figure:
\begin{itemize}
\item $(\nu, m)=(-1,3)$ for  Slow-Roll Inflation,
\item $(\nu, m)=(2,3)$ for Wands's model,
\item $(\nu, m)=(2,0)$ for the CSTB cosmos.
\end{itemize}
The CST bounce model is an antithesis to the slow roll  inflation scenario
 while the original matter-bounce model by Wands fails to fall on the curves of time independence~\footnote{%
This was  called ``duality" in~\cite{Li:2013bha}. In this paper we reserve the use
of ``duality'' for the strong/weak duality arisen  in  the AdS/CFT correspondence.}.

\begin{figure}[h!]
\centering
\includegraphics[width=0.8\textwidth]{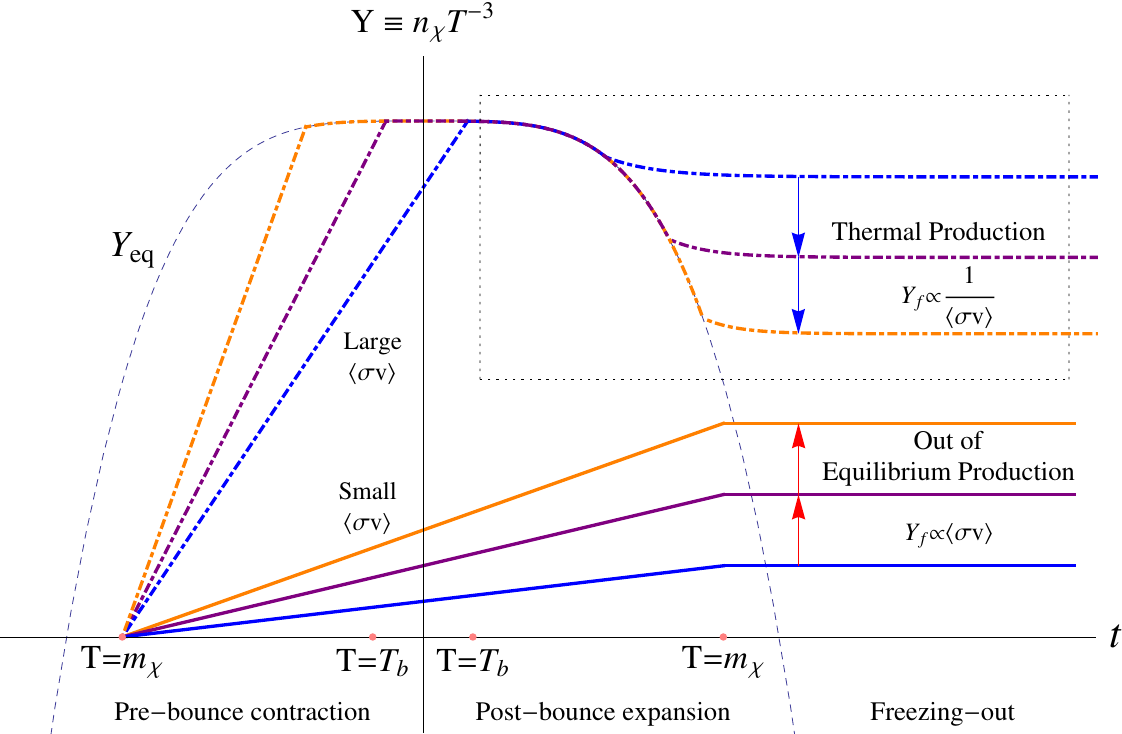}
\caption{%
The dark matter production and evolution in time is plotted. The Y-axis is dark matter
number density per entropy. A few characteristic temperatures are indicated on the time axis. When dark matter is produced in plenty abundance to reach the thermal equilibrium envelope, early universe information is washed out. When dark matter production exceed
its co-annihilation rate, the observed dark matter relic density constraint then relates dark matter mass and its coupling strength~\cite{Li:2014era}. }\label{fig:relic}
\end{figure}

With  the success of the primordial density  generation
 we proceeded to obtain testable predictions
from the bounce universe and their experimental verifications.
In a model independent way, we studied the dark matter generation and
evolutionary dynamics in a bounce universe~\cite{Li:2014era, Cheung:2014nxi}.
We found that when dark matter coupling was of the WIMP order, dark matter
 was produced in plenty abundance and  attained   thermal equilibrium.
In this case there was no constraint on the mass of dark matter produced.
However when the dark matter coupling was not of the WIMP level, but much
weaker instead, dark matter was produced in an out-of-thermal-equilibrium route~\footnote{Dark matter is produced, but the decay  process of dark matter is negligible.};
and thus the early universe information was encoded in the thermal evolution
of the background universe as  depicted in Fig.~\ref{fig:relic}.
Relic abundance constraint then puts dark matter
mass and coupling strength on a characteristic curve~\cite{Li:2014era}
which could be tested in dark matter detection experiments~\cite{Cheung:2016wik}.

We then proceeded to predict the number of dark
matter particles detected per kilogram-year for heavy dark matter detections in nuclear-recoil experiments~\cite{Cheung:2014pea} and for
 light dark matter in electron-scattering experiments~\cite{Vergados:2016niz}.
Together with cosmological surveys~\cite{Cai:2014bea} we are optimistic that
 in the near future the bounce universe can be distinguished
 from inflationary  scenario.

In this  paper we shall focus on the evolution of the spectral index of the
 primordial perturbations  as the  CSTB  cosmos  goes through a  bounce.
 We recall that the spectrum of density perturbations  is generated
during  the  matter dominated contraction, and we would like to know whether
 its scale invariance as well as its time dependence are altered
 by the bounce process, noting  that at the bounce point  the size of the universe
 reaches its minimum and the gravitational interactions may be strongly
 coupled. We  therefore have no prior knowledge  whether the bounce dynamics
 is  under control.

To the merit of the coupled-scalar-tachyon bounce universe model, it is
a model constructed  by taking a consistent truncation of  D-branes dynamics.
It is then straightforward to embed the model in type IIB string theory
and use the AdS/CFT correspondence to study the evolution of the primordial
density fluctuations across the bounce point when  gravity becomes strongly coupled.
Since the bulk/boundary correspondence is also a  strong/weak duality,
it means that we can map the fluctuations onto the boundary prior to the bounce point when the bulk dynamics is well under control;
and let this dynamics be
monitored by the boundary gauge theory during the bounce process, when the
gauge fields on the boundary are weakly coupled. After the bounce we can map back to the bulk and extract the information of
the scale invariance when the gravity sector is becoming weakly coupled
again. These references~\cite{Brandenberger:2016egn, Ferreira:2016gfg, Deger:2016axo, Das:2016eao, Chatterjee:2016bhj} inspire our work as well as many  aspects of our  method.

The few salient features of the CST bounce universe should be incorporated
  at the beginning of the study.  We need to make the cosmic background time dependent with a 4-dim Friedmann-Robertson-Walker  being a cross section inside the $AdS_{5}$ space.
The coupled Scalar-Tachyon  bounce universe
avoids  the violation of the null energy condition by having a spatial $S^{3}$~\cite{Borde:1993xh, Borde:1996pt} and obeying the ``soft bounce''
condition~\cite{Borde:1993xh}.
How to set up the analysis will be covered in Section~\ref{sec:dilaton}.
In particular  we present a time dependent dilaton solution with nonzero Ramond-Ramond charges in Type IIB string theory.
We describe the cosmic background in which CSTB model can be embedded.
In Section~\ref{sec:gauge-fields},  the equations of motion governing  the boundary gauge fields are  solved near the bounce point and  matched up
at different evolutionary phases.
Finally the spectrum can be studied in detail  throughout  the entire bounce process.
In Section~\ref{sec:disc}, we summarise and discuss  the results of the investigation.  We conclude  with an  outlook for  further studies on
 early universe models.

%
\section{A time dependent dilaton solution to Type IIB supergravity}
\label{sec:dilaton}

In string theory, a  Dp-brane  is an object with $p$ spatial dimensions
on which fundamental strings can end.
Randall and Sundrum¡«\cite{Randall:1999vf, Randall:1999ee} found a warped solution with two stacks of D3-branes in (3,1) spacetime separated in
the fifth direction in  an $AdS_5$.
The standard model fields live on the visible brane  and the other hidden
brane  can be moved to infinite distance.
 Thus  we can view our four dimensional bounce universe to be a D3-brane~\footnote{%
 To be more specific this pair  of D3-branes  should be an end product of
 non-BPS D4-branes and  anti-D4-branes  annihilation in Type IIB string vacuum.
 Lacking a dynamic description of the D-brane annihilation process for cosmological modelling we shall be restricting our attention on the resultant D3-branes themselves.}
and embed it into a full $AdS_5 \otimes S^5$ spacetime, and set up the AdS/CFT dynamics thereof.

To achieve our goal we  incorporate and develop  on many pioneering attempts of
applying   AdS/CFT correspondence~\cite{Maldacena:1997re} to  cosmological
studies~\cite{Kumar:2015gln, Bzowski:2015clm, Kumar:2015jxa,
Barbon:2015ria, Engelhardt:2015gla, Heidenreich:2015wga,
Engelhardt:2015gta, Enciso:2015qva,Banerjee:2015fua, Battarra:2014tga, Engelhardt:2014mea, Morrison:2014jha, Brandenberger:2013zea,
Enciso:2013lza, Smolkin:2012er, Enciso:2012wu, Barbon:2011ta,
Awad:2009bh, Awad:2008jf, Craps:2008cj, Awad:2007fj, Turok:2007ry,
Chu:2007um, Das:2006dz, Chu:2006pa, Hamilton:2005ju, Hertog:2005hu,
Durrer:2002jn}.
Recently  in~\cite{Brandenberger:2016egn} a specific  recipe is provided
for  determining  the scale dependence in the cosmological perturbations
when a cosmic singularity is encountered in the bulk spacetime.
We also find these works~\cite{Ferreira:2016gfg, Deger:2016axo,
Das:2016eao, Chatterjee:2016bhj} enlightening in the course of our investigation.

The  AdS/CFT correspondence is  also  a strong/weak duality: when the bulk
fields are strongly coupled, the boundary is described by a weakly coupled
field theory, and vice versa.
 The  bulk fields  have dual  operators  prescribed by the boundary theory.
 The dilaton  field  is related to the energy density of the gauge fields,
  and the gauge coupling of the boundary theory is determined by the vacuum expectation values  of the  dilaton.
 Therefore  the first step  is to find a time dependent solution of the
  dilaton  equation which, in turn, determines  the dynamics of  gauge fields  on the boundary.
When the boundary gauge field theory becomes weakly coupled
during the contraction, we can map the density perturbations in the bulk
 onto the boundary and follow  their  evolution  through the bounce holographically  using the dynamics of the gauge fields.

The  bounce process in the
bulk could be potentially violent or highly singular in nature. Luckily
this is not the case for the CST bounce  model. On the one hand it enjoys a string
theoretical completion at high energy and on the other hand the CST bounce universe
 has a minimum radius bigger than the Planck length, $\frac{1}{a_{min}^2} = \frac{8\pi}{3}  G_N V_0 = \frac{4{G_N}}{3g_{s}} {(\frac{m_{s}}{2\pi})^{4}}$.
Combining these facts one can conclude that the gauge fields on the boundary evolve smoothly as the bulk space undergoes a bounce.
After the bounce, we map the evolved gauge field fluctuations -- using again
the AdS/CFT dictionary -- back to the bulk. We can do this when the gravitational  dynamics
returns  to  a  weakly coupled state. The operation described above thus allows us to compare the post-bounce spectrum  with the pre-bounce spectrum. We can
check whether the scale invariance of the spectrum is respected by
the bounce process, or not.

To this end we need to  generalise the AdS/CFT correspondence to incorporate
time dependence of the model in order to  study how  the
spectrum of primordial  density perturbations,
which is generated prior to  the bounce in CSTB model,
is evolved through the bounce to our observable universe.
To this end we  need to find a time dependent solution  to the
dilaton equation  in Type IIB supergravity with nonzero Ramond-Ramond potentials~\cite{Bergshoeff:2001pv} which allows for
an $AdS_{5}\otimes S^{5}$ compactification.
The action of low energy effective theory of Type IIB string is given by~\cite{Polchinski:1998rr}:
\begin{eqnarray}
\label{eq:IIBaction}
\begin{split}S_{IIB}&=S_{NS}+S_R+S_{CS}\\
S_{NS}&=\frac1{2\kappa_{10}^2}\int d^{10}x\sqrt{-g}e^{-2\phi}\left(R+4\partial_\mu\phi\partial^\mu\phi-\frac1{12}\left|H_3\right|^2\right)\\
S_R&=-\frac1{4\kappa_{10}^2}\int d^{10}x\sqrt{-g}\left(\left|F_1\right|^2+\frac1{3!}\left|{\widetilde F}_3\right|^2+\frac1{2\times5!}\left|{\widetilde F}_5\right|^2\right)\\
S_{CS}&=-\frac1{4\kappa_{10}^2}\int C_4\wedge H_3\wedge F_3\end{split}~,\end{eqnarray}
where the form field strengths are defined as
${\widetilde F}_3=F_3-C_0\wedge H_3$,
${\widetilde F}_5=F_5-\frac12C_2\wedge H_3+\frac12B_2\wedge F_3$,
and
$F_1=dC_0$, $F_3=dC_2$, $F_5=dC_4$, $H_3=dB_2$.
The Ramond-Ramond 5-form  fluxes, sourced by the D3-branes wrapped on the $S^{5}$, should be  self-dual:
${\widetilde F}_5=\ast{\widetilde F}_5$~\footnote{%
The field equations derived from the action~(\ref{eq:IIBaction})
 should be consistent with the self-duality but do not imply it.}.
The Ramond-Ramond fluxes  modify  the covariant constant spinor condition which allows  for $AdS_{5}\otimes S^{5}$ compactification of Type IIB string.
The lower forms Ramond-Ramond  gauge potentials couple to lower dimensional
D-branes.  Dilaton $\phi$ and its dynamics are thereafter the
  focal point of our study.

The deformed  $AdS_5\otimes S^5$  spacetime metric  with which we will be  working  is,
 \begin{equation} \label{eq:ads5s5}
ds^2=\frac{L^2}{z^2} \left[- dt^2+a^2(t)
  (\frac{dr^2}{1-r^2}+r^2(d\theta^2+\sin^2\theta d\varphi^2))
     +dz^2\right]  +  L^2d\Omega_5^2,
\end{equation}
where $d\Omega_5^2$ being the metric of the unit $S^{5}$,  and
 $a(t)$  being  the scale factor of the 4-dimensional closed FLRW
 universe  and $L$  the AdS radius.

The equations  of motion obtained by varying~(\ref{eq:IIBaction})
are~\cite{Sfetsos:2010uq}:
\begin{eqnarray}\label{2.3} \nonumber
&&R_{\mu\nu}+2\partial_\mu\partial_\nu\phi
   -\frac14{\left(H_3^2\right)}_{\mu\nu} \\
&=&e^{2\phi}\left[\frac12{\left(F_1^2\right)}_{\mu\nu}
   +\frac14{\left({\widetilde F}_3^2\right)}_{\mu\nu}
   +\frac1{96}{\left({\widetilde F}_5^2\right)}_{\mu\nu}
    -\frac14g_{\mu\nu}\left(F_1^2+\frac16{\widetilde F}_3^2
   +\frac1{240}{\widetilde F}_5^2\right)\right],
   \end{eqnarray}
\begin{eqnarray}
\label{2.4}
R-4\partial_\mu\phi\partial^\mu\phi+4\partial_\mu\partial^\mu\phi
-\frac1{12}H^2&=&0~, \\
\label{2.5}
\ast{\widetilde F}_3\wedge H_3+d\ast dC_0&=&0~, \\
 \label{2.6}
2d\ast{\widetilde F}_3+H_3\wedge{\widetilde F}_5+\frac12B_2\wedge d {\widetilde F}_5-d C_4\wedge H_3&=&0~, \\
d \label{2.7}\ast{\widetilde F}_5 - H_3\wedge F_3 &=&0~, \\
\label{2.8}
-2d(e^{-2\phi}\ast H) + 2d(C_0\ast{\widetilde F}_3)
+  dC_2\wedge{\widetilde F}_5  +  \frac12C_2\wedge
d{\widetilde F}_5-dC_4\wedge dC_2 &=& 0~,
\end{eqnarray}
where   $\mu,\nu=0,1...9$
and the subscripts $p$ denote the ranks of the Ramond-Ramond fields.

We need to make some sensible assumptions as well as reasonable
simplifications  in order to solve  this formidable  array of equations.
A common  ansatz   for  self-duality  ${\widetilde F}_5={ F}_5$
is~\cite{Macpherson:2014eza}:
\begin{equation} \label{eq:derf}
\begin{array}{l}
\begin{aligned}
   {\widetilde F}_5
=& r(\sqrt{-\det g_{\alpha\beta}}
   dx^0\wedge dx^1\wedge dx^2\wedge dx^3\wedge dx^4  \\
&-\sqrt{\det g_{ab}}
    dx^5\wedge dx^6\wedge dx^7\wedge dx^8\wedge dx^9)~,
\end{aligned}\\
\end{array}
\end{equation}
with  $r$ being  a constant related to the total fluxes piercing through the $S^{5}$.
Coordinates on $AdS_5$ and $S^5$ are labelled by
$\alpha,\beta=0,...4$ and $a,b=5,...9$  respectively.
Recalling  that
 ${\widetilde F}_5=dC_4-\frac12C_2\wedge dB_2+\frac12B_2\wedge dC_2$,
we can further assume that $B_2$ and $C_2$ live in the $AdS_5$  and
$dC_4$ lives in the $S^5$.
In the orthonormal basis we express them as:
\begin{equation}
B_2=f_1dy^0\wedge dy^i+f_2dy^i\wedge dy^j
   +f_3dy^i\wedge dy^4+f_4dy^0\wedge dy^4~,
\end{equation}
\begin{equation}
C_2 = g_1dy^0\wedge dy^i+g_2dy^i\wedge dy^j
   +g_3dy^i\wedge dy^4+g_4dy^0\wedge dy^4~,
\end{equation}
where $i=1,2,3$, $\{dy^\mu\}$ are the orthonormal basis,
i.e. $dy^\mu=\sqrt{g_{\mu\mu}}dx^\mu$.

To  ameliorate  difficulty caused by the form fields
we assume that the coefficients $f_1 \cdots$, $g_1 \cdots$
are at most linear in  $y^0$ and $y^4$.
Then  the $AdS_5$ part of ${\widetilde F}_5$:
\begin{equation}
\begin{array}{l}
\begin{aligned}
\frac12(B_2\wedge dC_2-C_2\wedge dB_2)
  = &\frac32\lbrack f_1\frac{\partial g_2}{\partial y^4}
      +f_3\frac{\partial g_2}{\partial y^0}
       +f_2(\frac{\partial g_1}{\partial y^4}
       +\frac{\partial g_3}{\partial y^0})\\
    &-g_1\frac{\partial f_2}{\partial y^4}
      -g_3\frac{\partial f_2}{\partial y^0}
       -g_2(\frac{\partial f_1}{\partial y^4}
        +\frac{\partial f_3}{\partial y^0})\rbrack~.
\end{aligned}\\
\end{array}
\end{equation}
We will henceforth  take these $f_{i}$ to be constants and $g_{j}$ to be
linear in  $y^0$ and $y^4$.
The  constant, $r$,
mentioned above~(\ref{eq:derf}) becomes:
\begin{equation}
r=\frac32( f_1h_3+f_3h_2+f_2h_1)~,
\end{equation}
where
$h_1=\frac{\partial g_1}{\partial y^4}+\frac{\partial g_3}{\partial y^0}$,
$h_2=\frac{\partial g_2}{\partial y^0}$ and
$h_3=\frac{\partial g_2}{\partial y^4}$.
Since $f_4$ and $g_4$ do  not  appear in the equations of the form fields,
we will  take them to be zero.  Thus we have,
\begin{equation}
H_3=dB_2=0~,
\end{equation}
\begin{equation}
dC_2=h_1dy^0  \wedge dy^i \wedge dy^4
      +  h_2dy^0 \wedge dy^i\wedge dy^j
    + h_3dy^i\wedge dy^j\wedge dy^4~,
\end{equation}
\begin{equation}
dC_4=-rdy^5\wedge dy^6\wedge dy^7\wedge dy^8\wedge dy^9~.
\end{equation}
Putting these ansatz of the tensor fields into equations
(\ref{2.5}) to (\ref{2.8}) we arrive at
\begin{equation} \label{2.17}
\frac{\partial C_0}{\partial y^4}=-r\frac{h_3}{h_1}~,
\end{equation}
\begin{equation}  \label{2.18}
\frac{\partial C_0}{\partial y^0}=-r\frac{h_2}{h_1}~,
\end{equation}
\begin{equation}  \label{2.19}
h_1^2=h_2^2-h_3^2~.
\end{equation}

Likewise it is obvious that  we  should endow  the axion field $C_0$
linear dependence  in
 $y^0$ and the spatial direction, $y^4$, transverse to
our 4-dimensional universe inside the $AdS_{5}$, to be
 consistent with
the homogeneous and isotropic nature of the 4-dim FRW spacetime.
So far we have set up the ansatz for the form  fields with two free
 coefficient functions  $f_{i}$  and $h_{j}$.

At this point we are ready to  tackle~(\ref{2.4}), the
Euler-Lagrange equation of $\phi$:
\begin{equation}  \label{eq:phieom}
2\Delta_\mu\partial_\nu\phi
=4\partial_\mu\phi\partial_\nu\phi
-\frac{1}{2}  g_{\mu\nu} ( R + 4 g^{\lambda\rho} \partial_{\lambda} \phi \partial_{\rho} \phi)~.
\end{equation}
Putting Equations~(\ref{2.17}) to~(\ref{eq:phieom}) into~(\ref{2.3}),
together with  the  metric~(\ref{eq:ads5s5}),
we obtain  the equations of $\phi$ when  $\mu\nu=00,ii,44$,
\begin{equation}
\displaystyle{\frac{3(\dot{a}^2+1)}{a^2}-\frac{6}{z^2} + 2\dot\phi^2
   +2\phi^2_{,z}+\frac{6}{a^2}\,\phi^2_{,i}}
=e^{2\phi}\,\frac{L^2}{z^2}(\frac{r^2h_2^2}{2h_1^2}
   -3h_1^2-3h_2^2+\frac92h_3^2)~,
\end{equation}
\begin{equation}
\frac{6a^2}{z^2}-2a\ddot a - \dot a^2-1  + 2a^2\dot\phi^2
   -2a^2\phi_{,z}^2-2\phi_{,i}^2
= \displaystyle{e^{2\phi}\, \frac{a^2L^2}{z^2}\, \frac{15h_1^2}{2}}~,
\end{equation}
\begin{equation}
 \frac{6}{z^2}-\frac{3{\displaystyle\ddot a}}{a}
  -\frac{3 ({\displaystyle\dot a}^2+1)}{a^2}
  + 2\dot\phi^2 + 2\phi_{,z}^2 - \frac6{a^2}\phi_{,i}^2
=\displaystyle{e^{2\phi}\, \frac{L^2}{z^2}
 (\frac{r^2h_3^2}{2h_1^2}+3h_1^2+\frac{9}{2}\,h_2^2\, -\, 3h_3^2)}~,
\end{equation}
where $\dot\phi=\frac{\partial\phi}{\partial t}$,
$\phi_{,z}=\frac{\partial\phi}{\partial z}$ and $\phi_{,i}
=\frac{\partial\phi}{\partial y^i}$.

These are quadratic first-order partial differential equations of
$\phi$.  Usually they are also hard to solve.
If  we  view them as linear
equations of $\dot\phi^2$, $\phi_{,z}^2$ and $\phi_{,i}^2$,
life becomes  easier:
\begin{equation}  \label{2.24}
\dot\phi^2 = \frac14e^{2\phi}\frac{L^2}{z^2}\left(\frac{r^2h_2^2}{3h_1^2}
  +\frac{r^2h_3^2}{6h_1^2}+\frac{13}2h_1^2-\frac12h_2^2+2h_3^2\right)
  +\frac{3\ddot a}{4a}-\frac1{z^2}~,
\end{equation}
\begin{equation}  \label{2.25}
\frac{2\phi_{,i}^2}{a^2}
=\frac16\left[e^{2\phi}\frac{L^2}{z^2}\left(\frac{r^2}2-\frac{27}2h_1^2\right)+\frac{12}{z^2}-\frac{6({\displaystyle\dot a}^2+1)}{a^2} -\frac{3\ddot a}a\right]~,
\end{equation}
\begin{equation}  \label{2.26}
\phi_{,z}^2=\frac14e^{2\phi}\frac{L^2}{z^2}\left(\frac{r^2h_2^2}{6h_1^2}+\frac{r^2h_3^2}{3h_1^2}-\frac{13}2h_1^2+2h_2^2-\frac12h_3^2\right)
+\frac1{z^2}~.
\end{equation}

Compatible with homogeneity and isotropy,  we  let $\phi$ to be spatially homogeneous.
The right side of equation (\ref{2.25}) being  zero implies that
\begin{equation}\label{2.27}
e^{2\phi}\frac{L^2}{z^2}
=(\frac{6(\dot a^2+1)}{a^2}+
\frac{3\ddot a}a-\frac{12}{z^2})(\frac{r^2}2-\frac{27}2h_1^2)^{-1}~.
\end{equation}
Substituting  this back  into (\ref{2.24}) we  obtain
\begin{equation}  \label{2.28}
\dot\phi=\frac12\sqrt{\frac{6m({\displaystyle\dot a}^2+1)}{a^2}+\frac{3(m+1){\displaystyle\ddot a}}a-\frac{12m+4}{z^2}}~,
\end{equation}
with
$$
m\, =\, (\frac{r^2}3+\frac{r^2h_3^2}{2h_1^2}+\frac32h_1^2 +\frac92h_2^2-3h_3^2)(\frac{r^2}2-\frac{27}2h_1^2)^{-1}~.
$$
This constant encodes  the effects of form fields
$C_2$, $B_2$, $C_0$ in  the dynamics equation of  the dilaton, $\phi$.

At this point we have set up  a consistent  flux compactification of Type IIB string theory on $AdS_{5}\otimes S^{5}$ with a proper time dependence in the 4-dimensional FRW subspace
 of the $AdS_{5}$.
This is thus a starting point for utilising  AdS/CFT correspondence in
a cosmological study.
The  D3-brane  on which our universe resides is viewed  as a constant  $z$ slice of the $AdS_{5}$,   and  its evolution is denoted by  the  scale factor $a$.
In Section 3, we are going to utilise AdS/CFT correspondence to study the scale invariance of the primordial spectrum of density perturbations throughout the bounce process.

\section{The evolution of the  gauge-field fluctuations }
\label{sec:gauge-fields}
The analysis in this section closely follow the general techniques provided
by~\cite{Brandenberger:2016egn} with a few marked differences  which will be explained below in detail.
We  focus on normalisable bulk modes since we are interested in the  evolution of the  linearised bulk fluctuations.
Note that in this project we are interested in the primordial matter density
perturbations generated by the quantum fluctuations of the scalar field.
In the CSTB model this role is played by the tachyon  field~\footnote{The effective
mass of the scalar  field representing the separation of the stack of non-BPS D4-branes
and anti-D4-branes is much heavier, it thus describes the evolution of
the underlying cosmos~\cite{Li:2013bha}.}.
The scale dependence of this primordial spectrum
can be tested against  the current CMB measurements with extreme high precision.

Let us emphacise  that  we do not take  this scalar in the bulk to be the
dilaton field like it was  done in~\cite{Brandenberger:2016egn}.
 The  detailed analysis goes through as the mathematical framework
 allows to use the equations of motion  to evolve the boundary theory.
 Using the AdS/CFT correspondence we can  evolve the  corresponding linear fluctuation of  the gauge field $A_\mu$ on the boundary of $AdS_{5}$.
The initial conditions for the gauge fields on the boundary are set by the primordial fluctuations of the scalar fields via the correspondence; and thus
the  evolution of the gauge fields is thus well-defined at all time  and under control.
The primordial curvature perturbations in  the CSTB model  are related to the spectrum of tachyon matter  perturbations  by a constant factor in the long wavelength limit~\cite{Li:2013bha}.
The curvature perturbations in 5-dimensional bulk can also be studied
within this framework. In~\cite{Ferreira:2016gfg} it was shown that a generalisation of the uniform curvature gauge to the 5-dimensional case allows
one to relate the curvature perturbations to scalar field perturbations, as usual.
A straightforward calculation shows that this is also  the  case  in our model:
to the order of interest, the  calculations performed on  the scalar field
can be related to the curvature perturbations in  the longitudinal  gauge.

The boundary gauge theory is described by $\mathcal{N} = 4$ SYM theory,
the gauge coupling of which is determined by the dilaton,
 $g_{YM} ^2=e^{\phi}$.  Thus at linear level the effect of
 dilaton on the gauge field fluctuations will be its time-dependence,
 which we have  analysed  in detail  in  Section 2.
We shall now turn our attention to the asymptotical properties of
gauge coupling in order  to check the applicability of AdS/CFT correspondence
to our bounce universe.

\subsection{The scale factor of CSTB and the applicability of AdS/CFT}

\begin{figure}
\centering
\includegraphics[width=0.8\textwidth]{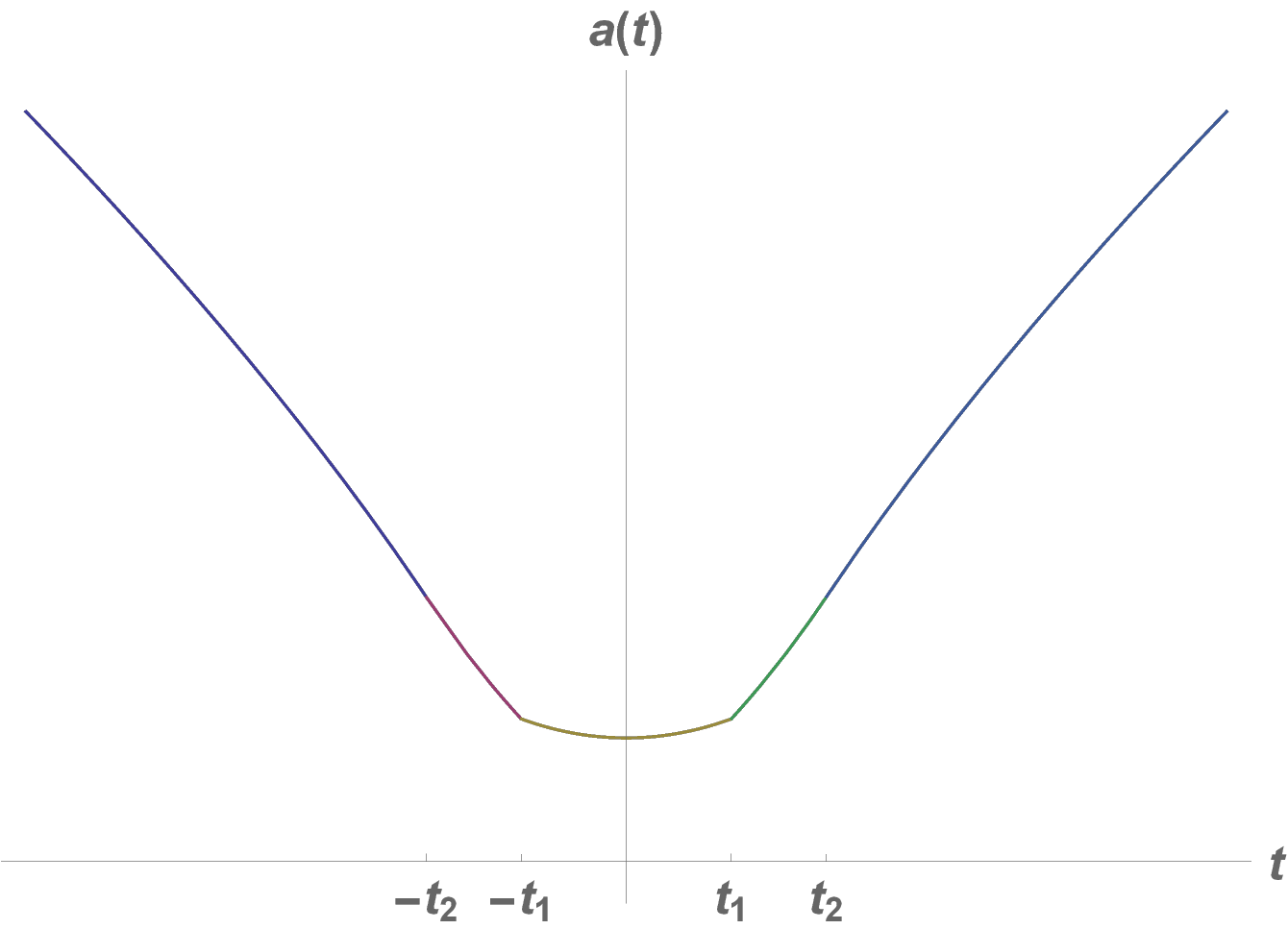}
\caption{The CSTB universe undergoes five phases: Tachyon matter dominated contraction, deflation, smooth bounce and locked inflation at the peak of the tachyon potential when the tachyon is locked by the fast oscillations of the
Higgs field. After that a  period of tachyon-matter-dominated rolling expansion ensues. }\label{fig:scalefactor}
\end{figure}

Recall that in the CSTB model, the universe undergoes a contraction phase,
deflation, the bounce, at which a phase of   ``locked inflation'' takes place
when the tachyon is locked at the peak of potential hill.
As the universe inflates, the oscillations of the Higgs fields get red-shifted
and eventually fails to lock the tachyon at the peak of its potential.
As  the tachyon rolls down its potential, it becomes a normal form of matter
with no  pressure~\cite{Sen:2002in}.
The cosmic evolution is schematically depicted in Figure~\ref{fig:scalefactor}.

In the subsequent calculations we are going to take three different,
but constant, values of the Hubble parameters~\cite{Cheung:2016oab, Li:2011nj},
\begin{equation}    \label{3.3.0}
{\rm Contraction}: a=a_2(-Ht)^{\frac{2}{3}},t\leq-t_2~,
\end{equation}
\begin{equation}\label{3.3}
{\rm Deflation}: a=a_1e^{-Ht},-t_2<t<-t_1~,
\end{equation}
\begin{equation}\label{3.3.1}
{\rm Smooth bounce}: a=a_0 {\rm cosh}{(Ht)}, -t_1\le t\le t_1~,
\end{equation}
\begin{equation}\label{3.3.2}
{\rm Locked inflation}: a=a_1e^{Ht},t_1<t<t_2~,
\end{equation}
\begin{equation}\label{3.3.3}
{\rm Expansion}: a=a_2(Ht)^{\frac{2}{3}},t\geq t_2~,
\end{equation}
where $t_1$ is the time when locked inflation starts,
$t_2$ is when it ends, $H$ is the Hubble constant during inflation and
$a_0$ is the minimal radius of the universe.
The bounce process is symmetric about  $t=0$~\footnote{%
In the absence of entropy creation the bounce process is symmetric
about  $t=0$. Considering entropy creation is outside the scope of this
paper and it does not affect the analysis undertaken in this work.}.

Now let us  check whether the correspondence is applicable at early and late times in our CSTB universe, insert~(\ref{3.3.3}) into~(\ref{2.27}) we obtain
\begin{equation}
e^{2\phi}=\frac{24-4z^2[t^{-2}
+3 a^{-2}_2 (Ht)^{-\frac{4}{3}}]}{L^2(27h_1^2-r^2)}~,
\end{equation}
the limit of 't Hooft coupling, $\lambda=g_s N=e^{\phi}N$, would be $\frac{N}{L}\sqrt{\frac{24}{27h_1^2-r^2}}$   when $|t|\rightarrow\infty$.
The framework we set up in Section 2  is fairly free: in the large $N$
 limit we  have an alternative ansatz to choose for  $h_i$ and $f_j$,
provided that  $27h_1^2-r^2$ is positive and large enough.
In this case the string coupling $g_s=e^{\phi}$ is small and  has a bound at all times,
which in turn makes the 't Hooft coupling $\lambda= g_s N$  decrease as $|t|$ decrease.
In other words the supergravity bulk theory
 will become less and less valid when the universe contracts.
In  the ensuing deflation phase~(\ref{3.3.2}), we have
\begin{equation}
e^{2\phi}=\frac{24-2z^2(9H^2+6a^{-2}_1e^{-2Ht})}{L^2(27h^2_1-r^2)}~.
\end{equation}
Similarly we can see that the gauge coupling $g_{YM}$ is small and decreases with $|t|$ decreasing, it is certainly bounded when $t_1<|t|<t_2$ .
Therefore the boundary gauge theory will become more and more valid as the universe  contracts.

During the smooth bounce process~(\ref{3.3.1}) the square of coupling is
\begin{equation}
e^{2\phi}=\frac{24-18(Hz)^2}{L^2(27h_1^2-r^2)}~,
\end{equation}
this is independent of time.
As discussed above, this choice of parameters can
give us a small coupling constant $g_{YM}^2$.
We note, however, that we may have to consider the effect of $z$.
It  cannot be too far from the boundary $z=0$, for instance if
$\frac{N}{L}\ll\sqrt{27h_1^2-r^2}$ then $z$ has to be
$Hz\sim \mathcal{O}(1)$.
In this case the Yang-Mills coupling and the  't Hooft coupling are both
small so the boundary gauge theory is well behaved  around the bounce point.
All in all  the AdS/CFT correspondence with appropriate choice of parameters
is an appropriate tool for studying the physics around the bounce point.

\subsection{An overview of the physics analysis around the bounce point}

As we shall see below we are going to map the bulk fluctuations onto the boundary.
 We would like to choose the mapping  point $\pm t_m$ to be when
$g_s N\sim \mathcal{O}(1)$, where the gauge theory becomes weakly coupled and  the gravity theory strongly coupled.
 And $|t_m|$ should be a little large than $|t_2|$, which means the mapping happens during contraction phase and is close to deflation. Besides, in the correspondence language $t_1$ should be the timing when the coupling decreases to and stays as a small enough constant.

The boundary theory is strongly coupled in the far past.
As the universe contracts, the bulk gravity theory becomes more and more strongly coupled. Long before  approaching  the bounce point,  we  map the
fluctuations  onto the boundary  where   it is  weakly coupled.
We monitor the evolution of the gauge fields  well after the bounce
 ends when the  bulk returns to a  weakly  coupled state.
In this and the following subsections we are going to use the notations
 in~\cite{Brandenberger:2016egn} and  review their setup
for the equation of motion for the gauge fields.

First making a gauge choice $A_0=0$ and imposing an additional constraint
 $\partial^iA_i=0$ so that the Gauss Law constraint is automatically solved.
By rescaling the gauge  fields,
$A_\mu \rightarrow {\widetilde A}_\mu\equiv e^{-\phi/2}A_\mu$,
an effective mass,
\begin{equation}  \label{3.2}
M^2_{YM}=\frac{\displaystyle\ddot \phi}{2}
-\frac{\displaystyle\dot \phi^2}{4}~,
\end{equation}
  appears  in the equations of motion for the rescaled gauge fields,
${\widetilde A}_i$,
\begin{equation}
-\partial^\mu \partial_\mu {\widetilde A}_i+M^2_{YM}{\widetilde A}_i=0~.
\end{equation}
Upon Fourier transformation,
\begin{equation}
{\widetilde A}_\nu(\xi^\mu)=\int^\infty_{-\infty}d^3\vec{k}c_A(\vec{k}){\widetilde A}_k(t)\epsilon_\nu e^{i\vec{k}\cdot\boldsymbol{\xi}}~,
\end{equation}
where $\epsilon_\nu$ is the unit polarization vector.
The  equations  of motion for the Fourier modes of the gauge fields
${\widetilde A}$ become~\cite{Awad:2008jf}:
\begin{equation}\label{3.1}
{\displaystyle\ddot {\widetilde A}}_k+(k^2+M^2_{YM}){\widetilde A}_k=0~.
\end{equation}

To summarise:  using the results in Section~2,   we  obtain  the scale
 factor,  $a$, and  the corresponding coupling constant,
 $g_s=g^2_{YM}=e^{\phi}$,
 in the five cosmic epochs of the CSTB universe.  In this section we  have  checked  thoroughly  the applicability of the AdS/CFT correspondence in the cosmological context of the CSTB model.  We hence conclude that, with a suitable choice of the ansatz parameters in Section~\ref{sec:dilaton}, the bulk theory can  be made weakly coupled  at early and late times.  The boundary theory can also be made  weakly coupled and well-behaved around the bounce point.
 This is what  we have realised  in the model building.
We briefly  review the setup for the gauge fields and their equations of motion.  We are interested to find out if the gauge fields acquire extra scale dependence due to the bounce dynamics.   Equipped with  the solution of dilaton  and the equations  of motion for ${\widetilde A}_k$, we can evolve gauge field  fluctuations at the boundary.

\subsection{The solution of ${\widetilde A}_k$ and its matching}
With the framework delineated above we are ready to  use the AdS/CFT
correspondence  to
 study the dynamics of the CSTB model in detail ---
by projecting  the bulk dynamics
onto the boundary and  evolve it using the boundary theory.
The mapping happens at contraction and expansion phases straddling the period when
the bulk becomes strongly coupled.
We shall  solve the equations  of motion in each phase:

\begin{enumerate}
\item{Contraction:}\\
Putting  (\ref{3.3.0}) into (\ref{2.28}) we obtain
\begin{equation}
\dot{\phi}=\frac{1}{2}\sqrt{\frac{6m}{a^2_2(Ht)^{\frac{4}{3}}}+\frac{2m-\frac{2}{3}}{t^2}-\frac{4(3m+1)}{z^2}}\approx\frac{1}{2}\sqrt{\frac{2m-\frac{2}{3}}{t^2}-\frac{4(3m+1)}{z^2}}~,
\end{equation}
here we ignore the $t^{-\frac{4}{3}}$ term in the root since $\frac{1}{a^2_2}$ is much smaller compared to $H^2$ at large $t$. Taking up to second order of $\frac{1}{t}$,
we have
\begin{equation}
\ddot{\phi}\approx\frac{1}{2}\sqrt{-\frac{4(3m+1)}{z^2}}[\frac{2m-\frac{2}{3}}{4(3m+1)}\frac{z^2}{t^3}+\frac{1}{2}(\frac{2m-\frac{2}{3}}{4(3m+1)})^2\frac{z^4}{t^5}]\approx0~,
\end{equation}
and
\begin{equation}
M^2_{YM}=-\frac{1}{16}(\frac{2m-\frac{2}{3}}{t^2}-\frac{4(3m+1)}{z^2})\equiv\frac{S}{t^2}+T~,
\end{equation}
where we donate the constants as $S\equiv\frac{\frac{2}{3}-2m}{16}$ and $T\equiv\frac{3m+1}{4z^2}$.
Putting it into (\ref{3.1}) yields
\begin{equation}
{\widetilde A}_k=L_1(k)\sqrt{-\beta_1t}J_{\sigma}(-\beta_1t)+L_2(k)\sqrt{-\beta_1t}N_{\sigma}(-\beta_1t)~,
\end{equation}
here $J_{\sigma}(t)$ and $N_{\sigma}(t)$ are two kinds of Bessel functions of order $\sigma\equiv\sqrt{\frac{1}{4}-S}$ and $\beta_1\equiv\sqrt{k^2+T}$, $L_{1,2}(k)$ are functions of $k$ and the notations are similar in the following.

\item{Deflation:}\\
Putting  (\ref{3.3}) and (\ref{2.28}) into (\ref{3.2}) we arrive at
\begin{equation} \label{eq:Mym}
M^2_{YM}=\frac{3mHe^{2Ht}}{2a^2_1\sqrt{(9m+3)H^2+\frac{6me^{2Ht}}{a^2_1}-\frac{12m+4}{z^2}}}-\frac{3}{16}(3m+1)H^2+\frac{3m+1}{4z^2}-\frac{3me^{2Ht}}{8a_1^2}~.
\end{equation}
In (\ref{eq:Mym}) near the matching point $t_1$, which is small, all the terms are effectively constant,
we  denote it as $M$.
 Putting it into (\ref{3.1}) yields
\begin{equation}
{\widetilde A}_k=D_1(k)e^{\beta_2 t}+D_2(k)e^{-\beta_2 t}~,
\end{equation}
where $\beta_2\equiv\sqrt{-k^2-M}$, $M\equiv\frac{3mH}{2a^2_1\sqrt{(9m+3)H^2+\frac{6m}{a^2_1}-\frac{12m+4}{z^2}}}-\frac{3}{16}(3m+1)H^2+\frac{3m+1}{4z^2}-\frac{3m}{8a_1^2}$~.

\item{The smooth bounce:} \\
Taking the first order of $t$ we obtain
\begin{equation} \label{eq:}
\begin{aligned}
M^2_{YM}&=\frac{3mH^4t}{\sqrt{-\frac{12m+4}{z^2}+3(m+1)H^2+\frac{6m}{a_0^2}}}
-\frac{1}{4}\left(-\frac{3m+1}{z^2}+\frac{3}{4}(m+1)H^2+\frac{6m}{a_0^2}\right)\\
&\equiv Pt+Q~,
\end{aligned}
\end{equation}
here $P\equiv\frac{3mH^4}{\sqrt{-\frac{12m+4}{z^2}+3(m+1)H^2+\frac{6m}{a_0^2}}}$ and $Q\equiv-\frac{1}{4}\left(-\frac{3m+1}{z^2}+\frac{3}{4}(m+1)H^2+\frac{6m}{a_0^2}\right)$ are constants, which yields
\begin{equation}
{\widetilde A}_k
=E_1(k){\rm Ai}\left[\frac{-k^2-Q-Pt}{(-P)^{\frac{2}{3}}}\right]
+E_2(k){\rm Bi}\left[\frac{-k^2-Q-Pt}{(-P)^{\frac{2}{3}}}\right]~,
\end{equation}
here $Ai[t]$ and $Bi[t]$ are Airy functions of $t$.

\item{Locked inflation:}
In this case everything is same as deflation except the value of~$t$.
Therefore
\begin{equation} \label{eq:3.10}
{\widetilde A}_k=F_1(k)e^{\beta_2 t}+F_2(k)e^{-\beta_2 t}~.
\end{equation}

\item{Expansion:}
In this case everything is same as contraction except the value of~$t$.
Therefore
\begin{equation}
{\widetilde A}_k=R_1(k)\sqrt{\beta_1t}J_{\sigma}(\beta_1t)+R_2(k)\sqrt{\beta_1t}N_{\sigma}(\beta_1t)~,
\end{equation}
where we re-iterate that $\{ \,D,\, E,\, F,\, R\}_{1,2}(k)$ are not constant
coefficients   but are assumed to be  functions of $k$.
\end{enumerate}

We take the asymptotic expansion of large $t$ for Bessel functions so that:
\begin{equation}
L_1(k)\sqrt{-\beta_1t}J_{\sigma}(-\beta_1t)=L_1(k)\sqrt{\frac{2}{\pi}}\cos(-\beta_1t-\frac{1+2\sigma}{4}\pi)~,
\end{equation}
\begin{equation}
L_2(k)\sqrt{-\beta_1t}N_{\sigma}(-\beta_1t)=L_2(k)\sqrt{\frac{2}{\pi}}\sin(-\beta_1t-\frac{1+2\sigma}{4}\pi)~.
\end{equation}
Matching ${\widetilde A}_k$ and its derivative at $t=-t_2$:
\begin{equation}
L_1(k)\sqrt{\frac{2}{\pi}}\cos(\beta_1t_2-\frac{1+2\sigma}{4}\pi)+L_2(k)\sqrt{\frac{2}{\pi}}\sin(\beta_1t_2-\frac{1+2\sigma}{4}\pi)=D_1(k)e^{-\beta_2 t_2}+D_2(k)e^{\beta_2 t_2}~,
\end{equation}
\begin{equation}
\sqrt{\frac{2}{\pi}}\frac{\beta_1}{\beta_2}[-L_1(k)\sin(\beta_1t_2-\frac{1+2\sigma}{4}\pi)+L_2(k)\cos(\beta_1t_2-\frac{1+2\sigma}{4}\pi)]=D_1(k)e^{-\beta_2 t_2}-D_2(k)e^{\beta_2 t_2}~,
\end{equation}
which yields
\begin{equation}\label{D_1}
\begin{aligned}
D_1(k)=&\frac{2}{\pi}\frac{e^{-\beta_2t_2}}{2\beta_2}\{[\beta_2\cos(\beta_2t_2-\frac{1+2\sigma}{4}\pi)-\beta_1\sin(\beta_2t_2-\frac{1+2\sigma}{4}\pi)]L_1(k)\\
&+[\beta_1\cos(\beta_2t_2-\frac{1+2\sigma}{4}\pi)+\beta_2\sin(\beta_2t_2-\frac{1+2\sigma}{4}\pi)]L_2(k)\}~,
\end{aligned}
\end{equation}
and
\begin{equation}\label{D_2}
\begin{aligned}
D_2(k)=&\frac{2}{\pi}\frac{e^{-\beta_2t_2}}{2\beta_2}\{[\beta_2\cos(\beta_2t_2-\frac{1+2\sigma}{4}\pi)+\beta_1\sin(\beta_2t_2-\frac{1+2\sigma}{4}\pi)]L_1(k)\\
&+[-\beta_1\cos(\beta_2t_2-\frac{1+2\sigma}{4}\pi)+\beta_2\sin(\beta_2t_2-\frac{1+2\sigma}{4}\pi)]L_2(k)\}~.
\end{aligned}
\end{equation}

We assume the arguments of both Airy functions to be  small and
 that the $(-P)^{\frac{1}{3}}t$ term dominate.
 We  can thus   expand the Airy functions asymptotically
 to  first power in $q\equiv\frac{-k^2-Q-Pt}{(-P)^{\frac{2}{3}}}$:
\begin{equation}
E_1(k){\rm Ai}\left(q\right)=\frac{\left(\frac{1}{3}\right)^{\frac{2}{3}}}{\Gamma \left(\frac{2}{3}\right)}E_1(k)~,
\end{equation}
\begin{equation}
E_2(k){\rm Bi}\left(q\right)
=\left[\frac{\left(\frac{1}{3}\right)^{\frac{1}{6}}}{\Gamma \left(\frac{2}{3}\right)}+\frac{\left(\frac{1}{3}\right)^{\frac{5}{6}}}{\Gamma \left(\frac{4}{3}\right)}q\right]E_2(k)~.
\end{equation}

Now we can match ${\widetilde A}_k$ and its derivatives at the end of
deflation and at the beginning of inflation,
 which we denote as $-t_1$ and $t_1$ respectively.
 Matching ${\widetilde A}_k$ yields:
\begin{equation} \label{3.14}
D_1(k)e^{-\beta_2 t_1}+D_2(k)e^{\beta_2 t_1}
=\frac{\left(\frac{1}{3}\right)^{\frac{2}{3}}}
    {\Gamma \left(\frac{2}{3}\right)}E_1(k)
+\left[\frac{\left(\frac{1}{3}\right)^{\frac{1}{6}}}
   {\Gamma \left(\frac{2}{3}\right)}
+\frac{\left(\frac{1}{3}\right)^{\frac{5}{6}}}
    {\Gamma \left(\frac{4}{3}\right)} q_1\right]E_2(k)~,
\end{equation}
\begin{equation}
F_1(k)e^{\beta_2 t_1}+F_2(k)e^{-\beta_2 t_1}
=\left[\frac{\left(\frac{1}{3}\right)^{\frac{2}{3}}}
   {\Gamma \left(\frac{2}{3}\right)}
  +\frac{\left(\frac{1}{3}\right)^{\frac{4}{3}}}
    {\Gamma \left(\frac{4}{3}\right)}q_2\right]E_1(k)
  +\left[\frac{\left(\frac{1}{3}\right)^{\frac{1}{6}}}
    {\Gamma \left(\frac{2}{3}\right)}
 -\frac{\left(\frac{1}{3}\right)^{\frac{5}{6}}}
   {\Gamma \left(\frac{4}{3}\right)}q_2\right]E_2(k)~,
\end{equation}
where $q_1\equiv q|_{t=-t_1}$ and $q_2\equiv q|_{t=t_1}$.

Matching ${\displaystyle \dot{\widetilde A}}_k$ yields:
\begin{equation}
D_1(k)\beta_2 e^{-\beta_2 t_1}-D_2(k)\beta_2 e^{\beta_2 t_1}
   =\frac{\left(\frac{1}{3}\right)^{\frac{5}{6}}}
    {\Gamma \left(\frac{4}{3}\right)}(-P)^{\frac{1}{3}}E_2(k)~,
\end{equation}
and
\begin{equation} \label{3.17}
F_1(k)\beta_2 e^{\beta_2 t_1}-F_2(k)e^{-\beta_2 t_1}
= \frac{\left(\frac{1}{3}\right)^{\frac{4}{3}}}
   {\Gamma \left(\frac{4}{3}\right)}P^{\frac{1}{3}}E_1(k)
  +\frac{\left(\frac{1}{3}\right)^{\frac{5}{6}}}
    {\Gamma \left(\frac{4}{3}\right)}(-P)^{\frac{1}{3}}~.
\end{equation}

Solving¡«(\ref{3.14}) to¡«(\ref{3.17}) we obtain the solutions to $F_{1}$ and $F_{2}$ as follows:
\begin{dmath}  \label{3.18}
F_1(k)\, =\, \frac{1}{6\beta_2\Gamma\left(\frac{4}{3}\right)(-P)^{\frac{1}{3}}e^{2\beta_2 t_1}}\left[D_1(k)I_1+D_2(k)e^{2\beta_2 t_1}I_2\right]~,
\end{dmath}
where
\begin{equation*}
\begin{split}
I_1=&-3^{\frac{1}{3}}\Gamma\left(\frac{2}{3}\right)(-P)^{\frac{2}{3}}
 +\beta_2(-P)^{\frac{1}{3}}\left[3^{\frac{1}{3}}\Gamma\left(\frac{2}{3}\right) (q_1+q_2)+9\Gamma\left(\frac{4}{3}\right)\right]\\
& -\beta_2^2\left[3^{\frac{1}{3}}\Gamma\left(\frac{2}{3}\right)
    q_1q_2+3\Gamma\left(\frac{4}{3}\right)(q_1+2q_2)\right]~,
\end{split}
\end{equation*}
and
\begin{equation*}
\begin{split}
I_2=&-3^{\frac{1}{3}}\Gamma\left(\frac{2}{3}\right)(-P)^{\frac{2}{3}}
+\beta_2(-P)^{\frac{1}{3}}\left[-3^{\frac{1}{3}}\Gamma\left(\frac{2}{3}\right)(q_1+q_2)-3\Gamma\left(\frac{4}{3}\right)\right]\\
&-\beta_2^2\left[3^{\frac{1}{3}}\Gamma\left(\frac{2}{3}\right)q_1q_2
+3\Gamma\left(\frac{4}{3}\right)(q_1+2q_2)\right]~;
\end{split}
\end{equation*}
while
\begin{dmath}
\label{3.19}
F_2(k)=\frac{1}{6\beta_2\Gamma\left(\frac{4}{3}\right)(-P)^{\frac{1}{3}}}\left[-D_1(k)J_1+D_2(k)e^{2\beta_2 t_1}J_2\right]~,
\end{dmath}
with $J_{1}$ and $J_{2}$ given by
\begin{equation*}
\begin{split}
J_1\, =\, &-3^{\frac{1}{3}}\Gamma\left(\frac{2}{3}\right)(-P)^{\frac{2}{3}}
  +\beta_2(-P)^{\frac{1}{3}}\left[3^{\frac{1}{3}}
    \Gamma\left(\frac{2}{3}\right)   (q_1-q_2)
  +3\Gamma\left(\frac{4}{3}\right)\right]\\
  &+\beta^2_2\left[3^{\frac{1}{3}}\Gamma\left(\frac{2}{3}\right)q_1q_2
  +3\Gamma\left(\frac{4}{3}\right)(q_1+2q_2)\right]~,
  \end{split}
\end{equation*}
and
\begin{equation*}
\begin{split}
J_2\, =\, &3^{\frac{1}{3}}\Gamma\left(\frac{2}{3}\right)(-P)^{\frac{2}{3}}
   +\beta_2(-P)^{\frac{1}{3}}\left[3^{\frac{1}{3}}
     \Gamma\left(\frac{2}{3}\right)(q_1+q_2)
     +9\Gamma\left(\frac{4}{3}\right)\right]\\
&-\beta_2^2\left[3^{\frac{1}{3}}\Gamma\left(\frac{2}{3}\right)q_1q_2
   +3\Gamma\left(\frac{4}{3}\right)(q_1+2q_2)\right]~.
\end{split}
\end{equation*}

By now we have matched the solutions of ${\widetilde A}_k$ at $t=-t_2$ and
$t=\pm t_1$; similar calculations  take place at  $t=t_2$.
In this way,  we  obtain a full   chain of relations
from $L_{1,2}(k)$ to $R_{1,2}(k)$,
which  encodes  all  the information concerning the  scale dependence in the
power spectrum.
It is then straightforward to judge if any of these functions pick up
extra  $k$ dependence  due to the bounce.

In particular we are interested in small wavenumber limit,  $kt\ll 1$,
when the modes of perturbations are outside the Hubble horizon.
Therefore, we have,
\begin{equation}
\beta_1=\sqrt{k^2+T}=\sqrt{k^2+\frac{4(3m+1)}{z^2}}\approx\sqrt{T}~,
\end{equation}
and
\begin{equation}
\label{3.20}
\beta_2=\sqrt{-k^2-M}\approx \sqrt{-M}~.
\end{equation}
In addition,  $Q\sim M$, we have
\begin{equation}\label{3.21}
q_1=\frac{-k^2-Q+Pt_1}{(-P)^{\frac{2}{3}}}
\approx\frac{-Q+Pt_1}{(-P)^{\frac{2}{3}}}~.
\end{equation}
Similar argument goes for $q_2$.
Henceforth   $\beta_{1,2}$ and $q_{1,2}$ are independent of $k$, with
 $e^{\beta_2t}$, $I_{1,2}$ and $J_{1,2}$ being  independent of $k$ when $kt_{1,2}\ll1$.

All in all one sees plainly  that  no extra $k$ dependence is introduced into  the
above relations by  the bounce process:
the spectral index is  {\it not} altered by the bounce  upon  comparing the
$k$ dependence in    $L_{1,2}(k)$ and  $R_{1,2}(k)$.
To conclude:  the scale invariance of the primordial density perturbations, if exists prior to the bounce,
 will hold  throughout the bounce process in the CSTB bounce universe as it
  cannot be affected by the bounce dynamics.

A few  comments on using AdS/CFT correspondence  to reconstruct of the spectrum from boundary data are made here.
As mentioned in the beginning of this section, the position of our universe brane cannot be too far away from the boundary. The field at a bulk point is
reconstructed  by integrating the boundary data against the boundary-to-bulk propagator, and the region is defined  by the null geodesics connecting the bulk point to the boundary~\cite{Brandenberger:2016egn, Hamilton:2005ju}.
Therefore the  reconstruction  of the information can be done in a small region
around the ``mapping point'' well before the gauge theory becomes strongly coupled again.
The low energy effective theory used to model the  bounce  universe is that of
 the Higgs fields  and the tachyon field, as well as their coupling term.
 Unlike other field theoretical   model,  it has a high energy completion
 as a string theory~\cite{Li:2011nj}: there is no singularity even as the scale
 factor achieves  its minimal value,
 $$
 \frac{1}{a_{min}^2} = \frac{8\pi}{3}  G_N V_0 = \frac{4{G_N}}{3g_{s}} {(\frac{m_{s}}{2\pi})^{4}}~,
 $$
  where $m_{s}$ is the string mass.
The qualification for our discussions may lie in the choice of ansatz and the effects of  different compactifications.
A complete analysis is beyond the scope of this paper,
to which we will  return in a future work.
In the current project we
are mostly  following (the spirit of) the recipe provided in~\cite{Brandenberger:2016egn}
for the  reconstruction of the bulk data  from boundary.
 We can conclude within
this framework the $k-$dependence  of the bulk fluctuations is completely
 determined  by the $k-$dependence of the  fluctuations of the gauge fields,
 ${\widetilde A}_k(t)$,   on the boundary.
 Thus   the evolution of the gauge  fluctuations
 will preserve the scale invariance of the primordial spectrum of density perturbations
 in the bulk  across the bounce point  in the CST bounce universe model.


\section{Conclusion and discussion}
\label{sec:disc}

In this paper we use the AdS/CFT correspondence to study whether the spectral
index of the  primordial perturbations is altered by the  bounce dynamics in the Coupled-Scalar-Tachyon Bounce Universe model.
Considering the string background  of CSTB model, we first embed the four dimensional bounce universe into an $AdS_5\otimes S^5$ background,
and find a time-dependent solution of dialton in type IIB  supergravity.
The dilaton solution is then used to  solve for  the dynamics of
gauge fields living on the boundary of the $AdS_{5}$.
We follow  the gauge fields through the  bounce process, as they are
weakly coupled, and match their solutions at different phases of cosmic evolution.

The combined profile of gauge field evolution is smooth across the bounce
point. The bounce process merely alters the amplitudes of different modes in the density perturbations, but it affects them in the same manner.
It therefore does not alter the intrinsic $k$ dependence in the
spectrum of primordial perturbations generated in the pre-bounce contraction.
We are led to  conclude that the scale invariance, if exists before the bounce, will hold throughout  the bounce process.
Nevertheless  as we can see from (\ref{D_1}), (\ref{D_2}), (\ref{3.18})
and (\ref{3.19}), as $k$ becomes larger and larger
 extra $k$ dependence begins to creep out in the spectral index,
the  implications of which are  under investigation.

No prior dilaton profile is assumed in our approach:
by finding an explicit solution to the dilaton equation of motion with
realistic  field  configurations of Type IIB superstring
on $AdS_{5}\otimes S^{5}$.
While a spacelike singularity exists in the bulk in the dilaton profile proposed earlier~\cite{Brandenberger:2016egn},
 there is no singularity in the bulk even at the time of bounce
 since the CSTB model has  a non-zero minimal radius at the bounce point.
 Furthermore we take the four dimensional bounce universe to be a closed
 FLRW  without violating NEC.
In~\cite{Brandenberger:2016egn}
  their result does not depend on the details of the time dependence of the boundary coupling. As  this is all about the evolution of boundary fluctuations through a potential singularity in the bulk spacetime,
 we are not sure whether their  result is applicable to general backgrounds
 and differently assumed dilaton profiles.

A clarifying remark on two different kinds of $k$-modes arisen in the
study is perhaps warranted. In many of the early universe models,
the Hubble parameter is no longer a constant.
Different $k$-modes exit and reenter the Hubble horizon at different times: the mode numbers pick up an implicit time
 dependence as a result of the time dependence in the  Hubble scale.
This makes  each  $k$-mode  in the primordial density perturbations
pick up an implicit time dependence:
only after this implicit time dependence is carefully taken into account
can the  time dependence in the spectrum be understood.
This is thus the key to the stability analysis on the spectrum
in  the CSTB model~\cite{Li:2013bha, Li:2012vi}.

The $k$-modes in (\ref{3.1}) are, however, the $k$-modes of the gauge
fields living on the boundary of the $AdS_{5}$.
They are  involved in the  mapping procedure  and merely encode
 the bulk dynamics holographically at  particular points on the boundary.
Therefore they  cannot inject or remove any time dependence in the
primordial spectrum.  Once the dynamics are  mapped  onto the boundary,
there is no more horizon crossing, the  gauge fields evolve under their own equations of motion. And the information of the $k$ dependence in the power spectrum, which is generated during contraction phase of bulk, is fully encoded in the solutions of gauge field fluctuations on the boundary.

We have made several assumptions and approximations  throughout
the analysis.  Different   solutions of the  dilaton could be obtained with different
 ansatz of  the Ramond-Ramond field configurations.
 We have simply chosen the most manageable, yet retaining the
 interesting physics,  configuration.
  With higher orders of time dependence in the dilaton field we
 have to expand $M^2_{YM}$ to the higher  order in $t$  when
 solving (\ref{3.1}).
 A systematic survey  of the field configurations and
 their  effects on the dilaton profile is beyond the scope of this paper.
 A dynamical mechanism, without prior assumptions on the  parameters,
 to select out the correct values of the parameters is highly desirable.
 These are  interesting physics, together with higher $\alpha\prime$ and
 higher string-loop effects
 on the string cosmological model, which we hope to address in future works.

Another line of research would be to properly set up and study the D-brane
and anti-D-brane annihilation process for cosmological modelling.
This constitutes the  (quantum) basis for building early universe model
in string theory.
This allows us to go beyond the effective field theory approach and
beyond kinematic analysis or  symmetry  arguments.
A  more realistic touch to string cosmology will eventually come from finding a
consistent string compactification that yields  a
nonsingular universe along with testable predictions that stand up to the
  the  array of precision cosmological  observations.

\section{Acknowledgments}

This research project has been supported in parts by the NSF China
under Contract No.~11775110, No.~11690034 and No.~11405084.
We also acknowledge the European Union's Horizon2020 research and innovation
programme (RISE) under the Marie Sk\'lodowska-Curie  grant agreement
No.~644121, and  the Priority Academic Program Development for
Jiangsu Higher Education Institutions (PAPD).


\addcontentsline{toc}{section}{References}

\end{document}